\documentclass[11pt, draftcls, onecolumn]{IEEEtran}
\usepackage{setspace}
\usepackage{multirow}
\usepackage[dvips]{graphicx}
\usepackage[cmex10]{amsmath}
\usepackage{amsmath}
\usepackage{amssymb}
\usepackage{amsfonts}
\DeclareGraphicsExtensions{ps,eps,pst,pdf}
\usepackage{cite}

\begin{document}

\title{Wideband Spectrum Sensing with Sub-Nyquist Sampling in Cognitive Radios}
\vspace{-2em}

\author{Hongjian~Sun,~\IEEEmembership{Member,~IEEE,}~Wei-Yu~Chiu,~\IEEEmembership{Member,~IEEE,} Jing Jiang,\\
A. Nallanathan,~\IEEEmembership{Senior Member,~IEEE,} and  H. Vincent Poor,~\IEEEmembership{Fellow,~IEEE}
\thanks{Copyright (c) 2012 IEEE. Personal use of this material is permitted. However, permission to use this material for any other purposes must be obtained from the IEEE by sending a request to pubs-permissions@ieee.org.}
\thanks{This manuscript has been accepted to be published in IEEE Transactions on Signal Processing. Manuscript received February 22, 2012; revised June 12, 2012; accepted July 21, 2012. The associate editor coordinating the review of this manuscript and approving it for publication was Prof. Shuguang Cui. Digital Object Identifier 10.1109/TSP.2012.2212892. }
\thanks{$^*$H. Sun and A. Nallanathan are with the Department of Electronic Engineering, King's College London, London, WC2R 2LS, UK. (Email: mrhjsun@hotmail.com; nallanathan@ieee.org)}
\thanks{W.-Y.~Chiu and H. V. Poor are with the Department of Electrical Engineering, Princeton University, Princeton, NJ 08544, US. (Email: chiuweiyu@gmail.com; poor@princeton.edu)}
\thanks{J. Jiang is with Center for Communication Systems Research, University of Surrey, UK. (Email: jing.jiang@surrey.ac.uk) }}

\maketitle

\vspace{-2em}

\begin{abstract}
Multi-rate asynchronous sub-Nyquist sampling (MASS) is proposed for wideband spectrum sensing. Corresponding spectral recovery conditions are derived and the probability of successful recovery is given. Compared to previous approaches, MASS offers lower sampling rate, and is an attractive approach for cognitive radio networks.
\end{abstract}


\section{Introduction}

The radio frequency (RF) spectrum is a limited natural resource, which is currently regulated by government agencies. The primary user (PU) of a particular spectral band has the exclusive right to use that band. Nowadays, on the one hand, the demands for the RF spectrum are constantly increasing due to the growth of wireless applications, but on the other hand, it has been reported that the spectrum utilization efficiency is extremely low. Cognitive radio (CR) is one of the promising solutions for addressing this spectral under-utilization problem~\cite{TVT1}. An essential requirement of CRs is that they must rapidly fill in spectrum holes (i.e., portions of the licensed but unused spectrum) without causing harmful interference to PUs. This task is enabled by spectrum sensing, which is defined as a technique for achieving awareness about the spectral opportunities and existence of PUs in a given geographical area \cite{poor1, poor2, magpoor}.

CR with a ``broader'' spectral awareness could potentially exploit more spectral opportunities and achieve greater capacity~\cite{GC2011}. Wideband spectrum sensing techniques therefore have attracted much attention in the research on CR networks. In~\cite{Tian2006}, a wavelet-based approach was studied for performing wideband spectrum sensing. It provides an advantage of flexibility in adapting to a dynamic wideband spectrum. Quan {\em et al.}~\cite{quan2, quan} proposed a multiband joint detection (MJD) approach for detecting the primary signal over multiple frequency bands. It has been shown that MJD performs well under practical conditions. In~\cite{bank1}, a filter-bank method was presented for sensing the wideband spectrum in a multicarrier communication system. The filter-bank system has been shown to have a higher spectral dynamic range than other traditional power spectrum estimation methods. Furthermore, compressive sensing (CS)~\cite{cs, RIP} technologies were introduced to implement wideband spectrum sensing in~\cite{ICC2012, scs1, wang, zeng, wide3, wide5}. Notably, these techniques take advantage of using sub-Nyquist sampling rates for signal acquisition, instead of the Nyquist rate, leading to reduced computational burden and memory requirements.

In this paper, we develop a multi-rate asynchronous sub-Nyquist sampling (MASS) system to perform wideband spectrum sensing. As the spectral occupancy is low, spectral aliasing (generated by sub-Nyquist sampling) is induced in each sampling branch to wrap the sparse spectrum occupancy map onto itself. The sensing requirements are therefore significantly reduced, i.e., the proposed MASS system has superior compression capability compared with the Nyquist sampling system. Considering the sub-Nyquist sampling in MASS, we then determine the recovery conditions under which the full wideband spectrum can be successfully reconstructed by using CS analysis. Compared with existing wideband spectrum sensing approaches, MASS has lower implementation complexity, higher energy efficiency, better data compression capability, and is more applicable to CR networks.

The rest of the paper is organized as follows. We briefly introduce the CS-based sensing scheme in Section~\ref{section2}. We then propose the MASS system in Section~\ref{section3}. Simulation results are presented in Section~\ref{section4}, followed by conclusions in Section~\ref{section5}.

\section{Problem Statement}
\label{section2}

In this paper, we assume that all CRs keep quiet during the spectrum sensing interval as enforced by protocols, e.g., at the medium access control (MAC) layer~\cite{quan2}. Therefore, the observed signal at a CR arises only from PUs and background noise. Suppose that the continuous-time signal $x_{\textrm{c}}(t)$ is received at a CR, and the frequency range of $x_{\textrm{c}}(t)$ is $0 \sim W$ (Hz). If the signal $x_{\textrm{c}}(t)$ is sampled at the sampling rate $f_{\textrm{s}}$ (Hz) for an observation time $T$, the sampled signal can be denoted by $x[n]=x_{\textrm{c}}(n/f_{\textrm{s}})$, $n=0, 1, \cdots, N-1$, (in vector form $\vec{x}\in \mathbb{C}^{N\times 1}$), where $N \stackrel{\triangle}{=} f_{\textrm{s}}T$ is assumed to be an integer. The discrete Fourier transform (DFT) spectrum of $\vec{x}$ can be calculated by $\vec{X}=\mathbf{F}\vec{x}$, where $\mathbf{F}$ denotes an $N$-by-$N$ DFT matrix $\mathbf{F}[h,n]\stackrel{\triangle}{=}e^{-j 2\pi nh/N}$ in which $j=\sqrt{-1}$. The difficulty arises when we consider the Shannon-Nyquist sampling theorem, which requires the sampling rate to be at least twice the bandwidth of the signal, i.e., $f_{\textrm{s}} \ge 2W$. On the one hand, we would like to realize ``broader'' spectral awareness at CRs (i.e., larger $W$), but on the other hand, the higher sampling rate will result in excessive memory requirements and prohibitive energy costs. This dilemma has motivated researchers to look for technologies to reduce the sampling rate $f_{\textrm{s}}$ while retaining $W$ by using CS theory~\cite{cs, RIP}.

\begin{figure}[t]
\centering
\centerline{\includegraphics[width=4.8in]{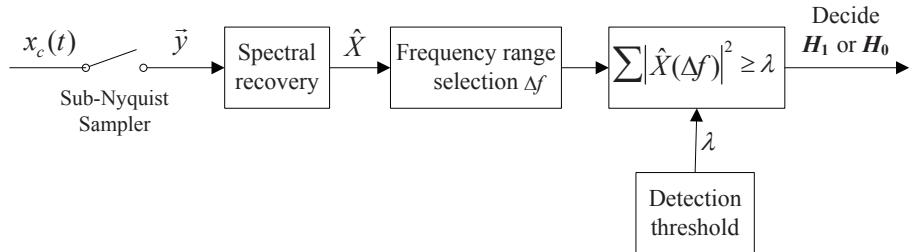}}
\vspace{-1.5em}
\caption{Diagram of CS-based spectrum sensing when using the spectral domain energy detection approach.}
\vspace{-1.5em}
\label{fig0}
\end{figure}

CS theory indicates that if the signal is sparse in a suitable basis, it can be exactly recovered from partial measurements. By ``sparse'' we mean that the signal can be represented using a few coefficients in some basis. Based on the fact of spectral sparseness~\cite{scs1}, it is reasonable to sample the signal $x_{\textrm{c}}(t)$ at a sub-Nyquist rate, reconstruct the spectrum $\vec{X}$ from partial measurements, and perform spectrum sensing using the reconstructed spectrum $\hat{X}$. A typical spectrum sensing approach is spectral domain energy detection~\cite{hongjian}. As depicted in Fig.~\ref{fig0}, this approach extracts the reconstructed spectrum in the frequency range of interest, e.g., $\Delta f$, and then calculates the signal energy in the spectral domain. The output energy is compared with a detection threshold (denoted by $\lambda$) to decide whether the corresponding frequency band is occupied or not, i.e., choosing between hypotheses $\mathcal{H}_{1}$ (presence of PUs) and $\mathcal{H}_{0}$ (absence of PUs). It is clear that the performance of sub-Nyquist-based spectrum sensing will highly depend on the recovery quality of the spectrum. In this paper, we will present a novel system, i.e., MASS, to sample the signal using sub-Nyquist sampling techniques, while enabling the spectrum $\vec{X}$ successfully recovered.

\section{Multi-rate Asynchronous Sub-Nyquist Sampling}
\label{section3}

\subsection{System and Signal Model}
\label{section3.1}

\begin{figure}[t]
\centering
\centerline{\includegraphics[width=5in]{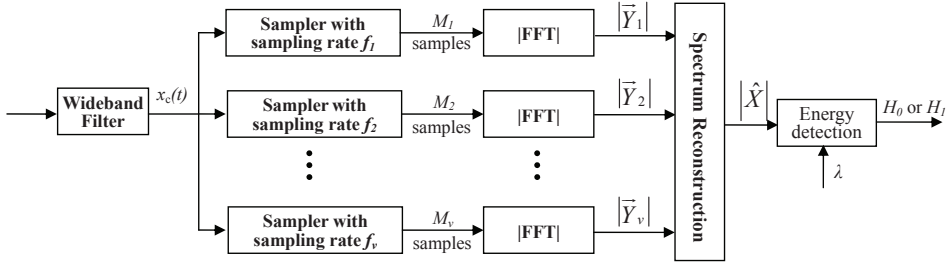}}
\vspace{-1.5em}
\caption{Schematic illustration of the multi-rate asynchronous sub-Nyquist sampling system in one CR node. The wideband filter has bandwidth $W$.}
\vspace{-1.6em}
\label{fig1}
\end{figure}

Suppose that a CR has $v$ sub-Nyquist sampling branches as shown in Fig.~\ref{fig1}. The wideband filter prior to the samplers removes frequencies outside the spectrum of interest, and is set to have bandwidth $W$. At the $i$-th branch, the low-rate sampler samples the received signal at the sub-Nyquist rate $f_i$ (Hz). In the observation time $T$ (second), the numbers of samples in these $v$ sampling branches are $M_{1}, M_{2}, \cdots, M_{v}$, respectively, where $M_i=f_iT~(\forall i \in [1, v])$. In addition, $M_{1}, M_{2}, \cdots, M_{v}$ are chosen to be different prime numbers that are of the order of $\sqrt{N}$, i.e., $M_{i}\sim \mathcal{O}(\sqrt{N})$, by controlling the sampling rate $f_i~(\forall i \in [1, v])$. The DFT spectrum of the sampled signal is then computed by applying the fast Fourier transform (FFT) to the samples in each branch. After that, these DFT spectra are used to reconstruct the wideband spectrum. We will employ an energy detection approach, and so we are interested in reconstructing only the magnitude of the spectrum, i.e.,~$|\vec{X}|$, resulting in the spectral magnitude estimate~$|\hat{X}|$.

Suppose that the received signal $x_{\textrm{c}}(t)$ is of finite support and absolutely summable. Using the sub-Nyquist rate $f_i<2 W$, we obtain the sampled signal $y_i[m]=x_{\textrm{c}}(m/f_i)=x_{\textrm{c}}(\frac{mT}{M_i})$, $m=0, \cdots, M_i-1$. The DFT spectrum of $\vec{y}_i$ is then calculated by $\vec{Y}_i=\mathbf{F}_s\vec{y}_i$, where $\mathbf{F}_s$ denotes the $M_i$-by-$M_i$ DFT matrix. The DFT spectrum of $\vec{y}_i$ is related to the continuous-time Fourier transform of $x_{\textrm{c}}(t)$ by~\cite{book1}
\begin{equation}
Y_i(f)=f_i \sum_{l=-\infty}^{\infty} X_{\textrm{c}}(f+ l f_i ) \label{shift}
\end{equation}
where $X_{\textrm{c}}(f)=\int_{-\infty}^{\infty}x_{\textrm{c}}(t)e^{-j2\pi t f}dt$ is the Fourier transform of $x_{\textrm{c}}(t)$. In other words, the DFT spectrum of the sampled data is a sum of the shifted Fourier transform spectrum of the continuous-time data.

Furthermore, if $x_{\textrm{c}}(t)$ is sampled at or above the Nyquist rate, i.e. $f_{\textrm{s}}=\frac{N}{T}\ge 2W$, the sampled signal can be written as $x[n]=x_{\textrm{c}}(n/f_{\textrm{s}})=x_{\textrm{c}}(\frac{nT}{N})$, $n=0, 1, \cdots, N-1$. The spectrum of $\vec{x}$ will be related to the continuous-time Fourier transform of $x_{\textrm{c}}(t)$ by $X(f)=f_{\textrm{s}} \sum_{l=-\infty}^{\infty} X_{\textrm{c}} (f+ l f_{\textrm{s}})$.
As the signal $x_{\textrm{c}}(t)$ is band-limited to $W$ and the sampling rate $f_{\textrm{s}}\ge 2W$, there will be no spectral aliasing phenomena in $X(f)$; thus, we can rewrite this relationship by $X(f)=f_{\textrm{s}} X_{\textrm{c}}(f)$, $\forall f\in \left[-\frac{W}{2},\frac{W}{2} \right]$. Because $X(f)$ has all information in $[-\frac{W}{2},\frac{W}{2}]$, we assume that $X(f)$ is zero everywhere except $f \in \left[-\frac{W}{2},\frac{W}{2} \right]$.  Substituting $X(f)=f_{\textrm{s}} X_{\textrm{c}}(f)$ into (\ref{shift}), we can obtain
\begin{equation}
Y_i(f)=\frac{f_i}{f_{\textrm{s}}} \sum_{l=-\infty}^{\infty} X(f+ l f_i), \;\;\; f+ l f_i \in \left[-\frac{W}{2},\frac{W}{2}\right]
\end{equation}
which has the discrete form of
\begin{eqnarray}
Y_{i}[m] &=& \frac{M_i}{N} \sum_{l=-\infty}^{\infty} X[ m+lM_i ], \;\;\; m+lM_i \in \left[-\left\lfloor \frac{N}{2} \right\rfloor, \left\lfloor \frac{N}{2} \right\rfloor \right]  \label{over}\\
&=&\frac{M_i}{N} \sum_{n=-\lfloor \frac{N}{2} \rfloor }^{\lfloor \frac{N}{2} \rfloor} X[ n ] \sum_{l=-\infty}^{\infty}\delta [n-(m+lM_i)], \;\;\; m \in \left[- \left\lfloor \frac{M_i}{2} \right\rfloor, \left\lfloor \frac{M_i}{2} \right\rfloor\right]
\label{sums3}
\end{eqnarray}
where $\lfloor a \rfloor$ is the floor function that gives the largest integer not greater than $a$, and $\delta[n]$ denotes the Dirac delta function that is zero everywhere except at the origin, where it is one. In matrix form, we write
\begin{equation}
\vec{Y}_{i}=\frac{M_i}{N}\mathbf{\Phi_i}\vec{X}
\label{cs}
\end{equation}
where the elements of $\mathbf{\Phi}_i \in \mathbb{R}^{M_i \times N}$ ($M_i < N$) can be represented by
$\mathbf{\Phi}_i\left[m+ \left\lfloor \frac{M_i}{2} \right\rfloor +1, n+ \left\lfloor \frac{N}{2} \right\rfloor+1 \right] = \sum_{l=-\infty}^{\infty} \delta \big[n -(m+lM_i) \big]$.
It can be seen that in each column of $\mathbf{\Phi}_i$, there is only one non-zero element and this has the value of one. In each row of $\mathbf{\Phi}_i$, the number of non-zero elements is at most $\lceil \frac{N}{M_i}\rceil$ (the ceiling function gives the smallest integer not less than $\frac{N}{M_i}$), which is the undersampling factor.

\subsection{Multi-rate Sub-Nyquist Sampling}
\label{section3.2}

Due to the sub-Nyquist sampling in each sampling branch, we should consider the effects of spectral aliasing. However, when the spectral sparsity $k \ll N$ and the sampling rate satisfies $M_i \sim \mathcal{O}(\sqrt{N})$, the probability of signal overlap is very small. The reader is referred to Appendix~A for the proof. In such a scenario, we concentrate on considering two cases: no signal at a particular value $m$ and one signal at $m$.
If there is no signal overlap, the following equation holds by using (\ref{cs}):
\begin{equation}
|\vec{Y}_{i}|=\left|\frac{M_i}{N}\mathbf{\Phi}_i \vec{X}\right| = \frac{M_i}{N} \mathbf{\Phi}_i |\vec{X}|
\label{why}
\end{equation}
where the last equation holds because the elements of $\mathbf{\Phi}_i$ are either zeros or ones, while each frequency bin of $\vec{Y}_{i}$ has no signal overlap from $\vec{X}$.

Furthermore, since all samplers observe the same spectral magnitude, i.e. $|\vec{X}|$, a concatenated equation relating $|\vec{X}|$ to the partial measurements can be formed by using (\ref{why}), i.e.,
\begin{eqnarray}
\vec{Y}\stackrel{\triangle}{=}\left(\begin{array}{c c}
\frac{N}{M_1}|\vec{Y}_{1}| \\
\frac{N}{M_2}|\vec{Y}_{2}| \\
\vdots \\
\frac{N}{M_v}|\vec{Y}_{v}|
\end{array}\right)
 = \left(\begin{array}{c c}
\mathbf{\Phi}_{1} \\
\mathbf{\Phi}_{2} \\
\vdots \\
\mathbf{\Phi}_{v}
\end{array}\right)
|\vec{X}|\stackrel{\triangle}{=}\mathbf{\Phi} |\vec{X}|
\label{noi}
\end{eqnarray}
where $\mathbf{\Phi}_{1},\mathbf{\Phi}_{2},\cdots,\mathbf{\Phi}_{v}$ are disjoint submatrices of $\mathbf{\Phi}\in \mathbb{R}^{(\sum_{i=1}^{v}M_{i}) \times N}$.

\emph{Remark 1:} The proposed system does not require exact synchronization between sub-Nyquist samplers. We merely require that the time offsets between sampling branches are sufficiently small so that the observed spectral magnitudes are quasi-stationary.  This is because, after the signal sampling and the spectral recovery, we employ the energy detection approach for performing spectrum sensing, and the use of energy detection only requires the spectral magnitude information. Due to this advantage, MASS can be applied to cooperative CR networks in which each CR requires one sub-Nyquist sampler and the spectral environments of different CRs are similar. In such scenarios, cooperative CRs are required to send the compressed magnitude information $|\vec{Y}_{1}|, \cdots, |\vec{Y}_{v}|$ to a fusion center for spectral recovery by using~(\ref{noi}).


To reconstruct the spectral magnitude (i.e., $|\vec{X}|$) using (\ref{noi}), we should carefully choose the measurement matrix $\mathbf{\Phi}$. We employ the mutual coherence~\cite{cs} to evaluate the suitability of $\mathbf{\Phi}$.

\emph{Definition 1:} Let $\mathbf{\Phi}$ be expressed as $\mathbf{\Phi}=[\vec{\phi}_{1} \; \vec{\phi}_{2} \cdots  \vec{\phi}_{N}]$, where $\vec{\phi}_{h}$ denotes the $h$-th column of the matrix $\mathbf{\Phi}$. Then the mutual coherence of the matrix $\mathbf{\Phi}$ is given by
\begin{equation}
\mu = \max_{l \neq z \in [1, N]} |<\hat{\phi}_{l},\hat{\phi}_{z}>|
\label{mu2}
\end{equation}
where $\hat{\phi}_{l}=\frac{\vec{\phi_{l}}}{\|\vec{\phi_{l}}\|_{2}}$ denotes the $\ell_{2}$ normalized column.

The aim is to keep $\mu$ to a minimum that allows the spectrum to be successfully recovered.
When the conditions in the following proposition are satisfied, the mutual coherence of $\mathbf{\Phi}$ in MASS can be determined by the number of sampling branches $v$.

\emph{Proposition 1:} Let $v$ samplers observe the spectral magnitude in the same observation time~$T$ and generate $v$ measurement vectors, i.e., $|\vec{Y}_{1}|, |\vec{Y}_{2}|,\cdots,|\vec{Y}_{v}|$. If the lengths of the measurement vectors are different primes, $M_{1}, M_{2}, \cdots, M_{v}$, that satisfy $M_{l}M_{z}>N, \; \forall l, z \in [1, v], l \neq z$, then the mutual coherence of the measurement matrix $\mathbf{\Phi}$ is given by
\begin{equation}
\mu=\max_{l \neq z} |<\hat{\phi}_{l},\hat{\phi}_{z}>|=\frac{1}{v}.
\label{mu}
\end{equation}

The proof of Proposition 1 is given in Appendix B.

\emph{Remark 2:} Donoho and Elad~\cite{optimal2} have proven that when the mutual coherence of the measurement matrix satisfies $\mu < \frac{1}{2k-1}$, the $k$-sparse signal can be successfully recovered. Thus, we know that if $\mu=\frac{1}{v}< \frac{1}{2k-1}$, i.e., $v>2k-1$, the spectral magnitude $|\vec{X}|$ can be exactly reconstructed. Proposition~1 illustrates that the number of samples in each sampler (or CR) is of the order of $\sqrt{N}$. When $M_i\sim \mathcal{O}(\sqrt{N})$, at least $2k$ sampling branches are required to recover the spectral magnitude.  To summarize, MASS needs the total number of measurements to be $\sum_{i=1}^vM_i\sim \mathcal{O}(k\sqrt{N})$.

Proposition 1 requires that the number of sampling branches $v>2k-1$ in order to reconstruct the spectrum. Actually, if the number of sampling branches $v$ is less than~$2k$, we could still reconstruct the spectrum with high probability. Next, we will investigate the probability of successful spectral reconstruction when we have an insufficient number of sampling branches, i.e., $v<2k$.

\emph{Proposition 2:} Let $v$ samplers observe the spectral magnitude in the same observation time $T$ and generate $v$ measurement vectors, i.e., $|\vec{Y}_{1}|, |\vec{Y}_{2}|,\cdots,|\vec{Y}_{v}|$. If the lengths of the measurement vectors are different primes and satisfy $M_{l}M_{z}>N, \; \forall l, z \in [1, v], l \neq z$, then the probability of successful spectral reconstruction is at least $1-\frac{2k-1}{v} \sum_{i=1}^v\frac{1}{M_i}$.

The proof of Proposition 2 is given in Appendix C.

\emph{Remark 3:} Proposition 2 shows that the probability of successful spectral reconstruction increases as either the number of sampling branches or the sub-Nyquist sampling rates increase. When $v=2k-1$, the probability of successful spectral reconstruction is at least $1-\sum_{i=1}^v\frac{1}{M_i}$. Further, as $M_i\sim \mathcal{O}(\sqrt{N})$ where $N=f_{\textrm{s}}T>2WT$, we know that, with ``broader" bandwidth~$W$, we could achieve higher probability of successful spectral reconstruction.

\section{Simulation Results}
\label{section4}

In simulations, we consider that there are $v$ sub-Nyquist sampling branches, each of which is equipped with a single low-rate sampler. In the $i$-th sampling branch, the received signal $x_{\textrm{c}}(t)$, as defined below, is sampled at a sub-Nyquist rate $f_i$ over an observation time $T$.
\begin{equation}
x_{\textrm{c}}(t)=\sum_{l=1}^{N_{b}}\sqrt{E_{l}}B_{l} \cdot \textrm{sinc}(B_{l}(t-\Delta))\cdot \cos(2\pi f_{l}(t-\Delta))+z(t)
\end{equation}
where sinc$(x)=\frac{\sin (\pi x)}{\pi x}$, $\Delta$ denotes a random time offset between sampling branches, and $z(t)$ is unit additive white Gaussian noise (AWGN). For the fading case, we generate $v$ different received powers $E_l$ according to the fading environment, and regenerate them for the next observation time. The received powers $E_l$ do not change within $T$, but vary randomly from branch to branch subject to the channel fading. The wideband signal $x_{\textrm{c}}(t)$ consists of $N_b=30$ non-overlapping subbands, whose bandwidths $B_l=0.5 \sim 5$~MHz, with carrier frequencies $f_{l} = 0 \sim 20$ GHz. Since the signal has a bandwidth of $W=20$~GHz, if it were sampled at the Nyquist rate for $T=2$~$\mu$s, the number of Nyquist samples would be $N=2WT=80,000$. However, in MASS we use $v=22$ sampling branches to sample the wideband signal with different sub-Nyquist rates, where $M_i\sim \mathcal{O}(\sqrt{N})$. Specifically, we select the first prime $M_1 \approx a\sqrt{N}$ ($a \ge 1$) and its $v-1$ neighboring and consecutive primes. The spectral magnitude $|\vec{X}|$ is reconstructed by applying the compressive sampling matching pursuit (CoSaMP) algorithm~\cite{cosamp} to (\ref{noi}). The spectral occupancy status is then decided by using the energy detection algorithm on the reconstructed spectrum.

\begin{figure}[t]
\begin{center}$
\begin{array}{cc}
\includegraphics[width=3in]{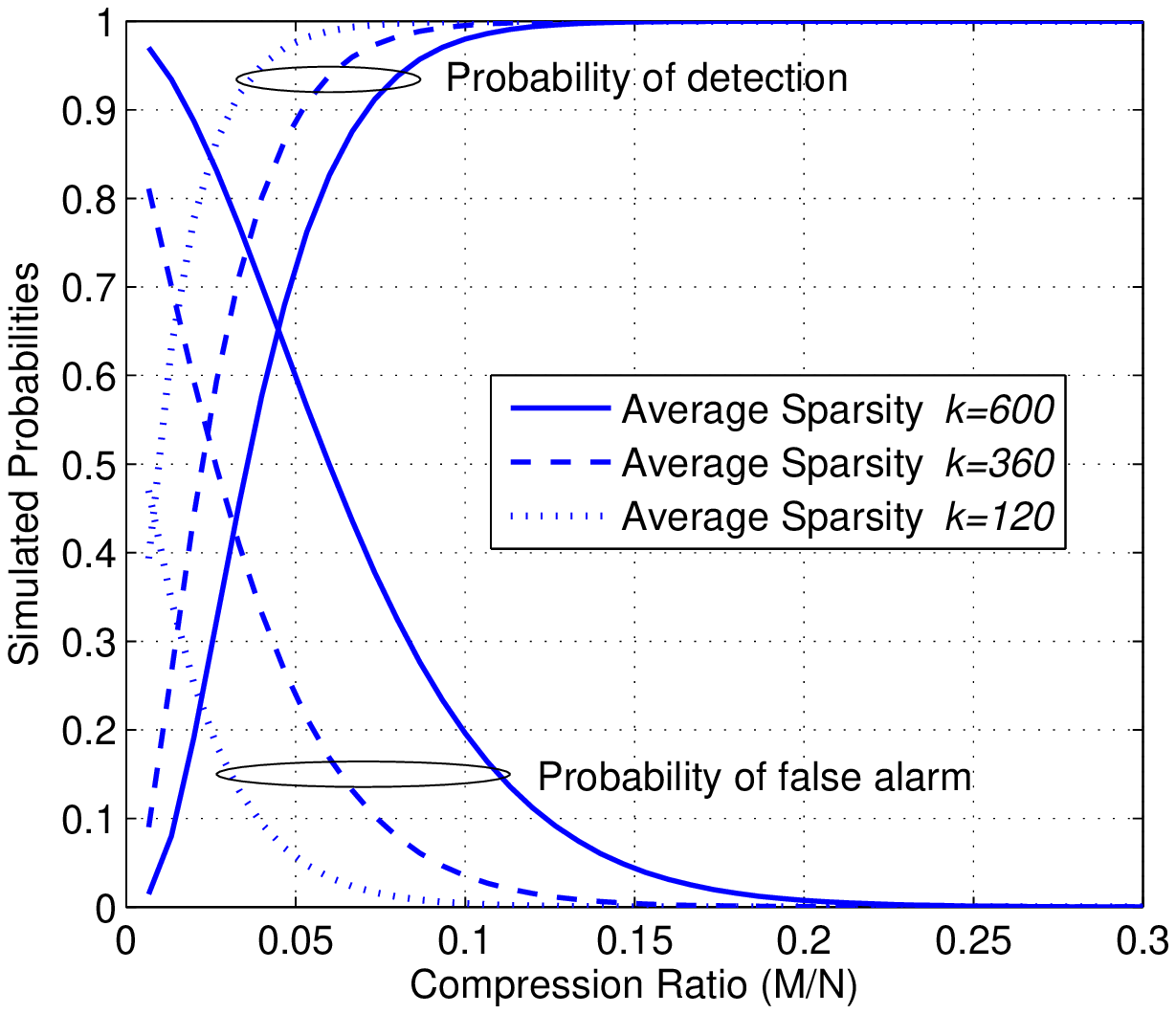} &
\includegraphics[width=3in]{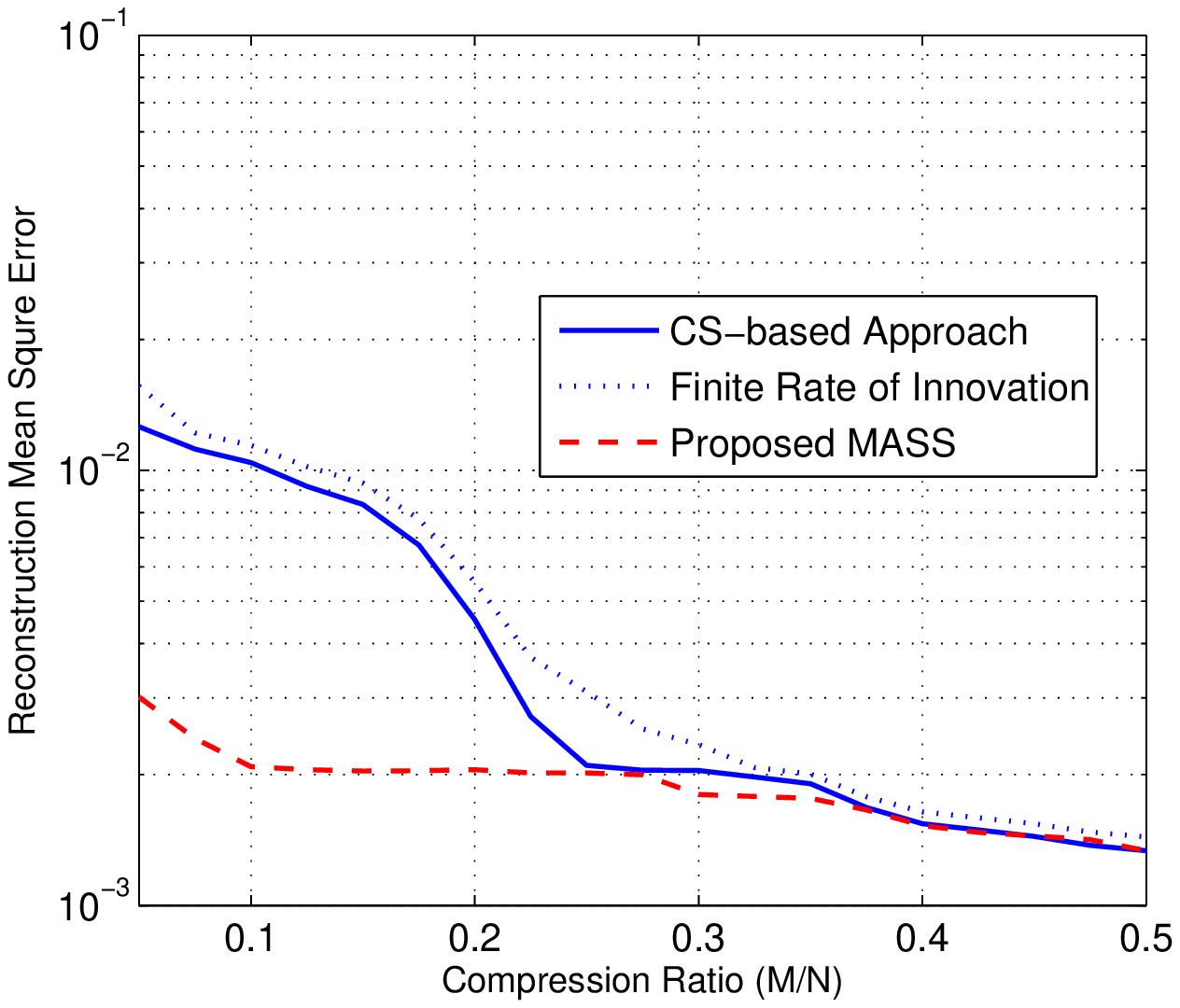}\\
(a) & (b)
\end{array}$
\end{center}
\vspace{-1.5em}
\caption{(a) Influence of the sparsity level and the compression ratio on the detection performance of MASS, when the signals from PUs suffer AWGN channels with average SNR$=10$ dB. (b) Comparison between the proposed system and the existing approaches when the compression ratio varies.}
\vspace{-1.5em}
\label{fig3}
\end{figure}

\begin{figure}[t]
\begin{center}$
\begin{array}{cc}
\includegraphics[width=2.9in]{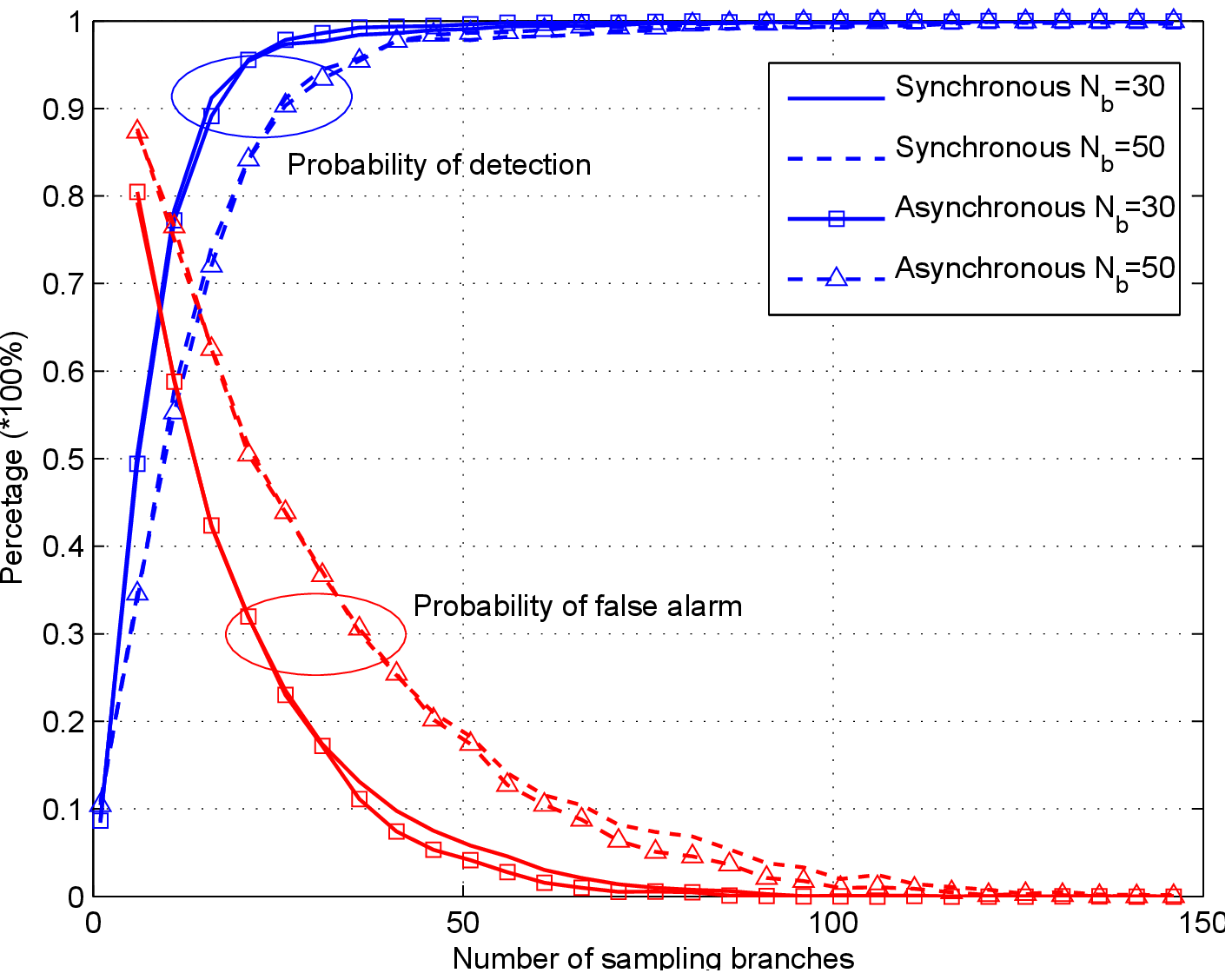} &
\includegraphics[width=2.9in]{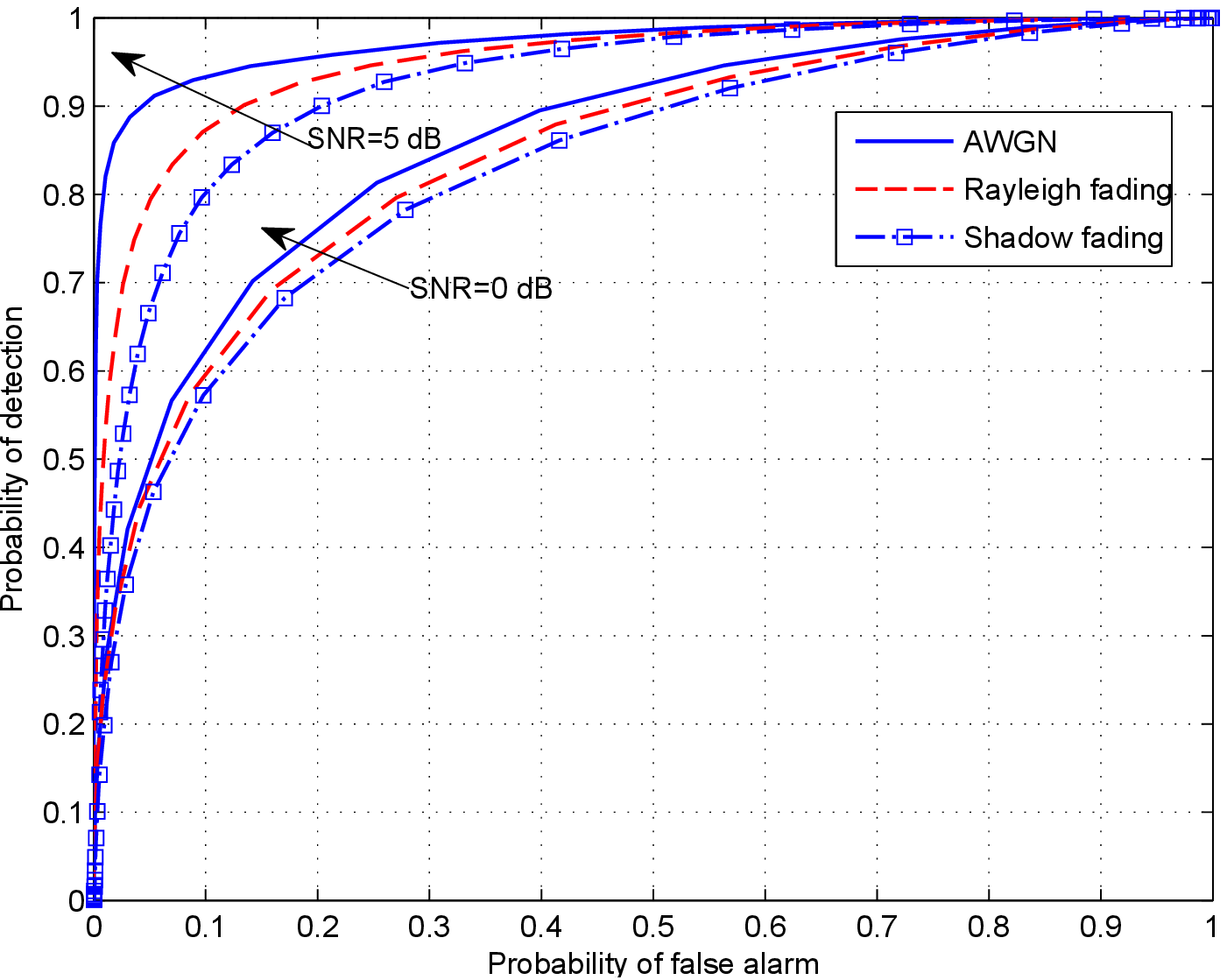}\\
(a) & (b)
\end{array}$
\end{center}
\vspace{-1.5em}
\caption{(a). Comparison of spectrum recovery performance for synchronous samplers and asynchronous samplers, with $N_b=30$ and $N_b=50$ over AWGN channels with average SNR$=10$ dB. (b). Performance of MASS over AWGN, Rayleigh, and shadow fading channels, with the number of sub-bands $N_b=30$, the compression ratio of $49.36\%$, and the standard deviation of shadow fading $5$~dB.}
\vspace{-1.6em}
\label{fig9}
\end{figure}

Fig.~\ref{fig3}(a) depicts the influence of the sparsity level $k$ and the compression ratio $\frac{M}{N}$  ($M\stackrel{\triangle}{=}\frac{\sum_{i=1}^v M_i}{v}$) on the detection performance of MASS. It illustrates that the lower the sparsity level, the better the detection performance. In addition, a higher compression ratio will lead to a lower probability of false alarm and a higher probability of detection. In Fig.~\ref{fig3}(b), we can see that the proposed system enables us to perform wideband spectrum sensing using fewer measurements (thus lower sampling rate). This is because, for the compression ratio below 0.2, the proposed system can achieve smaller reconstruction mean square error (MSE) when compared with both the CS-based approach and the finite rate of innovation approach.
Fig.~\ref{fig9}(a) shows the effect of imperfect synchronization between sampling branches. Compared with a reference clock, the asynchronous sampling branches have time offsets in the range of $0\sim0.8$ $\mu$s, while the total observation time is $2$ $\mu$s. It is evident that the detection performance of the asynchronous samplers is roughly the same as that of the synchronous samplers.
Fig.~\ref{fig9}(a) also illustrates that with more sampling branches, better spectrum sensing performance can be achieved. This is because with more sampling branches, a higher probability of successful spectrum recovery can be obtained.

In view of Remark 1 and the MASS's robustness against the imperfect synchronization in Fig.~\ref{fig9}(a), the proposed system could also be applied to CR networks in which each CR only has one sampling branch. In such a scenario, different branches could face independent and identically distributed (i.i.d.) fading channels.  In Fig.~\ref{fig9}(b), we analyze the detection performance of MASS over AWGN, Rayleigh, and shadow fading channels. When the signal-to-noise ratio (SNR) is zero decibels (dB), the performance of MASS over fading channels is roughly the same as that over non-fading AWGN channels. This is because the strength of the signal is mostly masked by the noise. By contrast, the detection performance of MASS over AWGN channels exceeds that over fading channels when SNR$=5$ dB. In addition, it is found that the performance of MASS over shadow fading channels is the poorest, in comparison to the case of AWGN and Rayleigh fading channels. Nonetheless, even over shadow fading channels, MASS has a probability of nearly $80\%$ of detecting the presence of PUs when the probability of false alarm is $10\%$, with the compression ratio $\frac{M}{N}=49.36\%$. On the contrary, Nyquist sampling systems, e.g., wavelet detection and multiband joint detection, must use at least $N$ measurements.

\begin{table}[!ht]
\centering
\tabcolsep 3pt
\footnotesize
\caption{Comparisons between wideband spectrum sensing techniques.} \vspace{-1.5em}
\begin{tabular}{c c c c}
\hline 
\raisebox{-2.5ex}{Technique} & \raisebox{-1ex}{Compression} & \raisebox{-1ex}{Sampling} & \raisebox{-1ex}{Implementation} \\ [0.5ex]
& \raisebox{1ex}{Capability} & \raisebox{1ex}{Scheme} & \raisebox{1ex}{Complexity} \\
\hline
Wavelet detection & no & Nyquist & low \vspace{0.6em}\\
Multiband joint detection & no & Nyquist & high \vspace{0.6em}\\
Filter-bank detection & no & Nyquist & high  \vspace{0.6em}\\ 
CS-based detection & yes & sub-Nyquist & medium \vspace{0.6em}\\
Proposed MASS & yes & sub-Nyquist & low \vspace{0.6em} \\ \hline
\end{tabular}\vspace{-1.5em}
\label{table1} 
\end{table}

In Table~\ref{table1}, we can see that, compared with Nyquist sensing techniques, MASS has outstanding compression capability due to multi-rate sub-Nyquist sampling. A similar system that has compression capability is the CS-based system. However, the state-of-the-art CS-based approaches require pseudo-random sequence generators as compression devices. In order to exploit spatial diversity in CR networks, each CR node should have a separate compression device, and transmit both compressed data and its measurement matrix to the fusion center. Even if measurement matrices are pre-stored, synchronization issues should still be considered because the un-synchronized measurement matrices could lead to false spectral recovery. In contrast, no separate device is required for generating the measurement matrix in MASS while the synchronization requirements are relaxed as shown in Fig.~\ref{fig9}(a). Thus, compared with the CS-based approaches, MASS has lower implementation complexity.

\section{Conclusions}
\label{section5}

In this paper, we have proposed a novel wideband spectrum sensing system, i.e., MASS. In this scheme, parallel low-rate samplers are used to sample the wideband signal at different sub-Nyquist rates. We have derived conditions for recovering the full spectrum by using CS analysis. The successful recovery probability has also been given. Simulation results have shown that MASS has superior compression capability compared with Nyquist sampling systems. Unlike other sub-Nyquist sampling systems, e.g., CS-based systems, MASS has been seen to be robust against lack of time synchronization and to have excellent performance in fading/shadowing scenarios. In summary, the proposed MASS not only has the benefits of low sampling rate, high energy-efficiency, and compression capability, but also is more amenable to implementation in CR networks in the presence of fading/shadowing.

\appendices

\section{Probability of Signal Overlap While Sub-Nyquist Sampling}

As the Nyquist spectrum is $k$-sparse, the probability of bin $n$ belonging to the spectral support is $P=\frac{k}{N}$. On letting $q$ denote the number of spectral components overlapped on bin $m$, the probability of no signal overlap can be given by using (\ref{over})-(\ref{cs}) as
\begin{equation}
\Pr(q < 2 ) = \Pr(q=0)\!+\!\Pr(q=1) = (1\!-\!P)^{\lceil \frac{N}{M_i} \rceil}\!+ \!\binom{\lceil \frac{N}{M_i} \rceil}{1} P (1\!-\!P)^{\lceil \frac{N}{M_i} \rceil\!-\!1}.
\label{jinsi}
\end{equation}
Substituting $P=\frac{k}{N}$ into (\ref{jinsi}) while using $M_i=\sqrt{N}$, we obtain
\begin{equation}
\Pr(q < 2 )  = (\frac{N\!-\!k}{N})^{\frac{N}{M_i}}\!+\! \frac{k}{M_i}(\frac{N\!-\!k}{N})^{\frac{N\!-\!M_i}{M_i}}  \!=  \!\frac{(\frac{N\!-k}{N})^{\sqrt{N}}(N\!-\!k\!+\!k\sqrt{N})}{N\!-\!k}.
\label{jinsi1}
\end{equation}
It can be tested that $\Pr(q < 2 )$ approaches 1 when $k \ll N$. Thus the probability of signal overlap approaches zero. $\Box$

\section{Proof of Proposition 1 }

Let $\hat{\phi}_n$ and $\hat{\phi}_h$ denote the $n$-th and the $h$-th ($h \neq n$) normalized column of the measurement matrix $\mathbf{\Phi}$, respectively. Using (\ref{cs}), they will have non-zero elements on the row indices $m_l=|n|_{\bmod (M_l)}$ and $g_l=|h|_{\bmod (M_l)}$ in the $l$-th sub-matrix, i.e., $\mathbf{\Phi}_l$. Obviously, $m_l$ could equal $g_l$ when $|h-n|$ is a multiple of $M_l$. To avoid this happening again in another sub-matrix, e.g., $\mathbf{\Phi}_z$, we could let $M_l$ and $M_z$ be different primes such that $M_lM_z>N$. This is because, if $M_l$ and $M_z$ are different primes and $M_lM_z>N$, $|h-n|$ cannot be a multiple of both $M_l$ and $M_z$. Thus, the maximum correlation of different columns in $\mathbf{\Phi}$ exists only when $m_l=g_l$ in a single sub-matrix $\mathbf{\Phi}_l$, in which case the mutual coherence $\mu=\frac{1}{v}$. $\Box$

\section{Proof of Proposition 2}

Based on Doob's maximal inequality~\cite{doob}, the following inequality holds:
\begin{equation}
\Pr(\mu>x)=\Pr\left(\max_{l \neq z} |<\hat{\phi}_{l},\hat{\phi}_{z}>|>x \right) \le \frac{\mathbb{E}\left[<\hat{\phi}_{l},\hat{\phi}_{z}>\right]}{x}.
\label{der}
\end{equation}
Let us define the event $\Omega_{l}$ such that $m_l$ equals $g_l$ as defined in the Appendix B, and $D_{l}=\frac{N}{M_{l}}$ to be the undersampling factor in the $l$-th sampling branch. The probability of the event $\Omega_{l}$ occurring is $\Pr(\Omega_{l})=\frac{\textrm{C}_{M_{l}}^{1}\textrm{C}_{D_{l}}^{2}}{\textrm{C}_{N}^{2}}=\frac{M_{l}D_{l}(D_{l}-1)}{N(N-1)}=\frac{N-M_{l}}{M_{l}(N-1)}<\frac{1}{M_{l}}$, where $\textrm{C}_n^k$ denotes the binomial coefficient. Then the expected value of $<\hat{\phi}_{j},\hat{\phi}_{h}>$ becomes
\begin{equation}
\mathbb{E}\left[<\hat{\phi}_{l},\hat{\phi}_{z}>\right]=\frac{1}{v} \sum_{i=1}^v\Pr(\Omega_{i})< \frac{1}{v} \sum_{i=1}^v\frac{1}{M_i}.
\label{expect}
\end{equation}
Replacing $x$ in (\ref{der}) by $x=\frac{1}{2k-1}$,  the following inequality can be obtained by using (\ref{der})-(\ref{expect}):
\begin{equation}
\Pr\left(\mu>\frac{1}{2k-1}\right) \le (2k-1)\mathbb{E}\left[<\hat{\phi}_{j},\hat{\phi}_{h}>\right]< \frac{2k-1}{v} \sum_{i=1}^v\frac{1}{M_i}.
\label{der1}
\end{equation}
Using the above inequality, we obtain
\begin{equation}
\Pr\left(\mu<\frac{1}{2k-1} \right)>1-\frac{2k-1}{v} \sum_{i=1}^v\frac{1}{M_i}.
\end{equation}
Considering the successful recovery condition in~\cite{optimal2}, i.e., $\mu<\frac{1}{2k-1}$, we complete the proof. $\Box$

\bibliographystyle{IEEEtran}
\bibliography{cwpf}

\begin{thebibliography}{10}
\providecommand{\url}[1]{#1}
\csname url@samestyle\endcsname
\providecommand{\newblock}{\relax}
\providecommand{\bibinfo}[2]{#2}
\providecommand{\BIBentrySTDinterwordspacing}{\spaceskip=0pt\relax}
\providecommand{\BIBentryALTinterwordstretchfactor}{4}
\providecommand{\BIBentryALTinterwordspacing}{\spaceskip=\fontdimen2\font plus
\BIBentryALTinterwordstretchfactor\fontdimen3\font minus
  \fontdimen4\font\relax}
\providecommand{\BIBforeignlanguage}[2]{{%
\expandafter\ifx\csname l@#1\endcsname\relax
\typeout{** WARNING: IEEEtran.bst: No hyphenation pattern has been}%
\typeout{** loaded for the language `#1'. Using the pattern for}%
\typeout{** the default language instead.}%
\else
\language=\csname l@#1\endcsname
\fi
#2}}
\providecommand{\BIBdecl}{\relax}
\BIBdecl

\bibitem{TVT1}
Y.-C. Liang, K.-C. Chen, G.~Y. Li, and P.~Mahonen, ``Cognitive radio networking
  and communications: An overview,'' \emph{IEEE Transactions on Vehicular
  Technology}, vol.~60, no.~7, pp. 3386--3407, Sept. 2011.

\bibitem{poor1}
S.~Chaudhari, V.~Koivunen, and H.~V. Poor, ``Autocorrelation-based
  decentralized sequential detection of {OFDM} signals in cognitive radios,''
  \emph{IEEE Transactions on Signal Processing}, vol.~57, no.~7, pp.
  2690--2700, July 2009.

\bibitem{poor2}
J.~Lunden, V.~Koivunen, A.~Huttunen, and H.~V. Poor, ``Collaborative
  cyclostationary spectrum sensing for cognitive radio systems,'' \emph{IEEE
  Transactions on Signal Processing}, vol.~57, no.~11, pp. 4182--4195, Nov.
  2009.

\bibitem{magpoor}
E.~Axell, G.~Leus, E.~G. Larsson, and H.~V. Poor, ``Spectrum sensing for
  cognitive radio: State-of-the-art and recent advances,'' \emph{IEEE Signal
  Processing Magazine}, vol.~29, no.~3, pp. 101--116, May 2012.

\bibitem{GC2011}
H.~Sun, A.~Nallanathan, J.~Jiang, D.~Laurenson, C.-X. Wang, and H.~Poor, ``A
  novel wideband spectrum sensing system for distributed cognitive radio
  networks,'' in \emph{Proc. IEEE Global Telecommunications Conference},
  Houston, TX, USA, Dec. 2011, pp. 1--6.

\bibitem{Tian2006}
Z.~Tian and G.~Giannakis, ``A wavelet approach to wideband spectrum sensing for
  cognitive radios,'' in \emph{Proc. IEEE Cognitive Radio Oriented Wireless
  Networks and Communications}, Mykonos Island, Greece, June 2006, pp. 1--5.

\bibitem{quan2}
Z.~Quan, S.~Cui, A.~H. Sayed, and H.~V. Poor, ``Optimal multiband joint
  detection for spectrum sensing in cognitive radio networks,'' \emph{IEEE
  Transactions on Signal Processing}, vol.~57, no.~3, pp. 1128--1140, Mar.
  2009.

\bibitem{quan}
------, ``Wideband spectrum sensing in cognitive radio networks,'' in
  \emph{Proc. IEEE International Conference on Communications}, Beijing, China,
  May 2008, pp. 901--906.

\bibitem{bank1}
B.~Farhang-Boroujeny, ``Filter bank spectrum sensing for cognitive radios,''
  \emph{IEEE Transactions on Signal Processing}, vol.~56, no.~5, pp. 1801
  --1811, 2008.

\bibitem{cs}
D.~Donoho, ``Compressed sensing,'' \emph{IEEE Transactions on Information
  Theory}, vol.~52, no.~4, pp. 1289--1306, April 2006.

\bibitem{RIP}
E.~Candes, J.~Romberg, and T.~Tao, ``Robust uncertainty principles: Exact
  signal reconstruction from highly incomplete frequency information,''
  \emph{IEEE Transactions on Information Theory}, vol.~52, no.~2, pp. 489--509,
  Feb. 2006.

\bibitem{ICC2012}
H.~Sun, A.~Nallanathan, J.~Jiang, and H.~V. Poor, ``Compressive autonomous
  sensing ({CASe}) for wideband spectrum sensing,'' in \emph{Proc. IEEE
  International Conferene on Communications}, Ottawa, Canada, June 2012, pp.
  5953--5957.

\bibitem{scs1}
Z.~Tian and G.~Giannakis, ``Compressed sensing for wideband cognitive radios,''
  in \emph{Proc. IEEE International Conference on Acoustics, Speech, and Signal
  Processing}, Honolulu, HI, USA, April 2007, pp. 1357--1360.

\bibitem{wang}
Y.~Polo, Y.~Wang, A.~Pandharipande, and G.~Leus, ``Compressive wide-band
  spectrum sensing,'' in \emph{Proc. IEEE International Conference on
  Acoustics, Speech, and Signal Processing}, Taipei, Taiwan, April 2009, pp.
  2337--2340.

\bibitem{zeng}
F.~Zeng, C.~Li, and Z.~Tian, ``Distributed compressive spectrum sensing in
  cooperative multihop cognitive networks,'' \emph{IEEE Journal of Selected
  Topics in Signal Processing}, vol.~5, no.~1, pp. 37--48, Feb. 2011.

\bibitem{wide3}
F.~Zeng, Z.~Tian, and C.~Li, ``Distributed compressive wideband spectrum
  sensing in cooperative multi-hop cognitive networks,'' in \emph{Proc. IEEE
  International Conference on Communications}, Cape Town, South Africa, May
  2010, pp. 1--5.

\bibitem{wide5}
Z.~Tian, E.~Blasch, W.~Li, G.~Chen, and X.~Li, ``Performance evaluation of
  distributed compressed wideband sensing for cognitive radio networks,'' in
  \emph{Proc. Int. Conf. Information Fusion}, Cologne, Germany, July 2008, pp.
  1--8.

\bibitem{hongjian}
H.~Sun, D.~Laurenson, and C.-X. Wang, ``Computationally tractable model of
  energy detection performance over slow fading channels,'' \emph{IEEE
  Communications Letters}, vol.~14, no.~10, pp. 924--926, Oct. 2010.

\bibitem{book1}
W.~A. Gardner, \emph{Statistical Spectral Analysis: A Nonprobabistic
  Theory}.\hskip 1em plus 0.5em minus 0.4em\relax Prentice-Hall, Inc., Upper
  Saddle River, NJ, USA, 1986.

\bibitem{optimal2}
D.~L. Donoho and M.~Elad, ``Optimally sparse representation in general
  (nonorthogonal) dictionaries via $\ell_1$ minimization,'' in \emph{Proc. Nat.
  Acad. Sci. USA}, vol. 100, no.~5, 2003, pp. 2197--2202.

\bibitem{cosamp}
D.~Needell and J.~Tropp, ``Co{S}a{MP}: Iterative signal recovery from
  incomplete and inaccurate samples,'' \emph{Applied and Computational Harmonic
  Analysis}, vol.~26, no.~3, pp. 301--321, 2009.

\bibitem{doob}
J.~L. Doob, \emph{Stochastic Processes}.\hskip 1em plus 0.5em minus 0.4em\relax
  London: Chapman \& Hall, 1953.

\end{thebibliography}

\end{document}